\documentclass[twoside,twocolumn,9pt]{article}
\usepackage{extsizes}
\usepackage[super,sort&compress,comma]{natbib} 
\usepackage[version=3]{mhchem}
\usepackage[left=1.5cm, right=1.5cm, top=1.785cm, bottom=2.0cm]{geometry}
\usepackage{balance}
\usepackage{mathptmx}
\usepackage{sectsty}
\usepackage{graphicx} 
\usepackage{lastpage}
\usepackage[format=plain,justification=justified,singlelinecheck=false,font={stretch=1.125,small,sf},labelfont=bf,labelsep=space]{caption}
\usepackage{float}
\usepackage{fancyhdr}
\usepackage{fnpos}
\usepackage{siunitx}
\usepackage{mhchem}
\usepackage[english]{babel}
\addto{\captionsenglish}{%
  
}
\usepackage{array}
\usepackage{droidsans}
\usepackage{charter}
\usepackage[T1]{fontenc}
\usepackage[usenames,dvipsnames]{xcolor}
\usepackage{setspace}
\usepackage[compact]{titlesec}
\usepackage{hyperref}
\usepackage{url}
\usepackage{color}
\usepackage{epstopdf}%This line makes .eps figures into .pdf - please comment out if not required.

\definecolor{cream}{RGB}{222,217,201}

\begin{document}

\pagestyle{fancy}
\thispagestyle{plain}
\fancypagestyle{plain}{
%%%HEADER%%%
\renewcommand{\headrulewidth}{0pt}
}
%%%END OF HEADER%%%

%%%PAGE SETUP - Please do not change any commands within this section%%%
\makeFNbottom
\makeatletter
\renewcommand\LARGE{\@setfontsize\LARGE{15pt}{17}}
\renewcommand\Large{\@setfontsize\Large{12pt}{14}}
\renewcommand\large{\@setfontsize\large{10pt}{12}}
\renewcommand\footnotesize{\@setfontsize\footnotesize{7pt}{10}}
\makeatother

\renewcommand{\thefootnote}{\fnsymbol{footnote}}
\renewcommand\footnoterule{\vspace*{1pt}% 
\color{cream}\hrule width 3.5in height 0.4pt \color{black}\vspace*{5pt}} 
\setcounter{secnumdepth}{5}

\makeatletter 
\renewcommand\@biblabel[1]{#1}            
\renewcommand\@makefntext[1]% 
{\noindent\makebox[0pt][r]{\@thefnmark\,}#1}
\makeatother 
\renewcommand{\figurename}{\small{Fig.}~}
\sectionfont{\sffamily\Large}
\subsectionfont{\normalsize}
\subsubsectionfont{\bf}
\setstretch{1.125} %In particular, please do not alter this line.
\setlength{\skip\footins}{0.8cm}
\setlength{\footnotesep}{0.25cm}
\setlength{\jot}{10pt}
\titlespacing*{\section}{0pt}{4pt}{4pt}
\titlespacing*{\subsection}{0pt}{15pt}{1pt}
%%%END OF PAGE SETUP%%%

%%%FOOTER%%%
\fancyfoot{}
\fancyfoot[LO,RE]{\vspace{-7.1pt}\includegraphics[height=9pt]{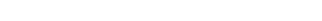}}
\fancyfoot[CO]{\vspace{-7.1pt}\hspace{13.2cm}\includegraphics{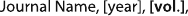}}
\fancyfoot[CE]{\vspace{-7.2pt}\hspace{-14.2cm}\includegraphics{head_foot/RF}}
\fancyfoot[RO]{\footnotesize{\sffamily{1--\pageref{LastPage} ~\textbar  \hspace{2pt}\thepage}}}
\fancyfoot[LE]{\footnotesize{\sffamily{\thepage~\textbar\hspace{3.45cm} 1--\pageref{LastPage}}}}
\fancyhead{}
\renewcommand{\headrulewidth}{0pt} 
\renewcommand{\footrulewidth}{0pt}
\setlength{\arrayrulewidth}{1pt}
\setlength{\columnsep}{6.5mm}
\setlength\bibsep{1pt}
%%%END OF FOOTER%%%

%%%FIGURE SETUP - please do not change any commands within this section%%%
\makeatletter 
\newlength{\figrulesep} 
\setlength{\figrulesep}{0.5\textfloatsep} 

\newcommand{\topfigrule}{\vspace*{-1pt}% 
\noindent{\color{cream}\rule[-\figrulesep]{\columnwidth}{1.5pt}} }

\newcommand{\botfigrule}{\vspace*{-2pt}% 
\noindent{\color{cream}\rule[\figrulesep]{\columnwidth}{1.5pt}} }

\newcommand{\dblfigrule}{\vspace*{-1pt}% 
\noindent{\color{cream}\rule[-\figrulesep]{\textwidth}{1.5pt}} }

\makeatother
%%%END OF FIGURE SETUP%%%

%%%TITLE, AUTHORS AND ABSTRACT%%%
\twocolumn[
  \begin{@twocolumnfalse}
{\includegraphics[height=30pt]{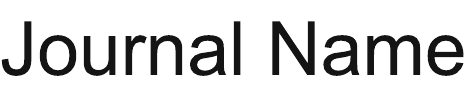}\hfill\raisebox{0pt}[0pt][0pt]{\includegraphics[height=55pt]{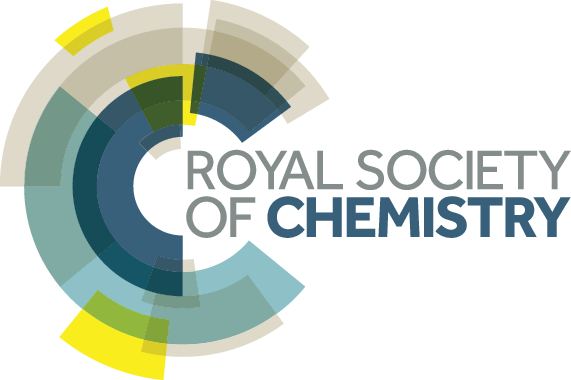}}\\[1ex]
\includegraphics[width=18.5cm]{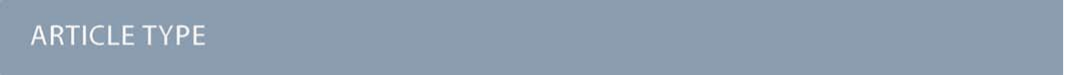}}\par
\vspace{1em}
\sffamily
\begin{tabular}{m{4.5cm} p{13.5cm} }

\includegraphics{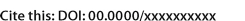} & \noindent\LARGE{\textbf{Lowering the Exponential Wall: Accelerating High-Entropy Alloy Catalysts Screening using Local Surface Energy Descriptors from Neural Network Potentials}} \\%Article title goes here instead of the text "This is the title"
\vspace{0.3cm} & \vspace{0.3cm} \\

 & \noindent\large{Tomoya Shiota,$^{\ast,\dagger,}$\textit{$^{a, b}$} Kenji Ishihara,\textit{$^{\dagger, b}$} and Wataru Mizukami\textit{$^{{\ast},a, b}$}} \\%Author names go here instead of "Full name", etc.

\includegraphics{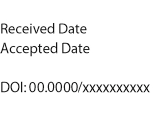} & \noindent\normalsize{Computational screening is indispensable for the efficient design of high-entropy alloys (HEAs), which hold considerable potential for catalytic applications. However, the chemical space of HEAs is exponentially vast with respect to the number of constituent elements, making even machine learning-based screening calculations time-intensive. To address this challenge, we propose a rapid method for predicting HEA properties using data from monometallic systems (or few-component alloys). Central to our approach is the newly introduced local surface energy (LSE) descriptor, which captures local surface reactivity at atomic resolution. We established a correlation between LSE and adsorption energies using monometallic systems. Using this correlation in a linear regression model, we successfully estimated molecular adsorption energies on HEAs with significantly higher accuracy than a conventional descriptor (i.e., generalized coordination numbers). Furthermore, we developed high-precision models by employing both classical and quantum machine learning. Our method enabled CO adsorption-energy calculations for 1000 quinary nanoparticles, comprising 201 atoms each, within a few days, considerably faster than density functional theory, which would require hundreds of years or neural network potentials, which would have taken hundreds of days. The proposed approach  accelerates the exploration of the vast HEA chemical space, facilitating the design of novel catalysts.} \\%The abstrast goes here instead of the text "The abstract should be..."

\end{tabular}

 \end{@twocolumnfalse} \vspace{0.6cm}

  ]
%%%END OF TITLE, AUTHORS AND ABSTRACT%%%

%%%FONT SETUP - please do not change any commands within this section
\renewcommand*\rmdefault{bch}\normalfont\upshape
\rmfamily
\section*{}
\vspace{-1cm}

%%%FOOTNOTES%%%

\footnotetext{\textit{$^{a}$Graduate School of Engineering Science, Osaka University, 1-3 Machikaneyama, Toyonaka, Osaka 560-8531, Japan; E-mail: shiota.tomoya.ss@gmail.com; mizukami.wataru.qiqb@osaka-u.ac.jp}}
\footnotetext{\textit{$^{b}$Center for Quantum Information and Quantum Biology, Osaka University, 1-2 Machikaneyama, Toyonaka 560-8531, Japan}}
\footnotetext{\textit{$^{\dagger}$These two authors contributed equally to this work.}}
\footnotetext{\textit{$^{\ast}$Corresponding auther}}
%Please use \dag to cite the ESI in the main text of the article.
%If you article does not have ESI please remove the the \dag symbol from the title and the footnotetext below.
%additional addresses can be cited as above using the lower-case letters, c, d, e... If all authors are from the same address, no letter is required

%%%END OF FOOTNOTES%%%

%%%MAIN TEXT%%%%

\section{\label{sec:level1}Introduction}

\begin{figure*}[h]
\centering
\includegraphics{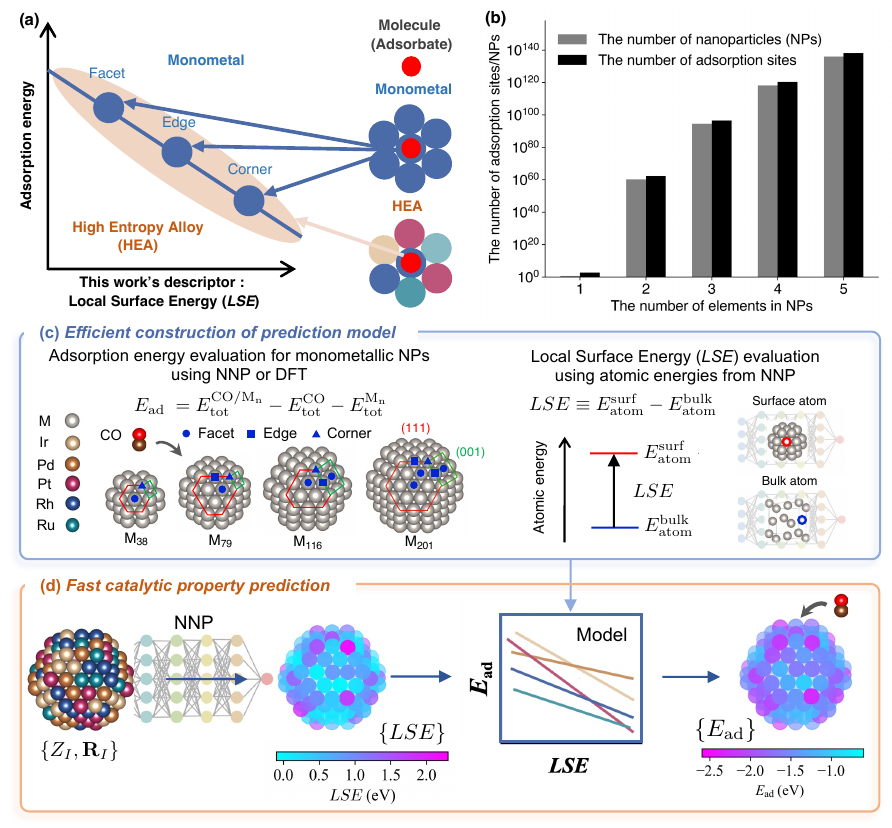}% Here is how to import EPS art
\caption{\label{fig:1} (a) Relationship between the molecular adsorption energy and the newly introduced descriptor, i.e., the local surface energy (LSE), for monometallic and high-entropy alloy (HEA) nanoparticle (NP) systems. The LSE descriptor developed in this study effectively captures the variation in adsorption energy across different adsorption sites. (b) Near-exponential increase in the number of NPs and adsorption sites as the number of elements in the NPs increases from monometallic to multicomponent systems.(c) The model is constructed through linear regression between the adsorption energies and the LSE using atomic energies from the NNP. The CO adsorption energies are evaluated on the ontop irreducible sites of facets (circles), edges (squares), and corners (triangles) of the monometallic NPs indicated by the blue symbols. (d) Fast catalytic property prediction workflow using neural network potentials (NNPs). The adsorption energy \({E_{\text{ad}}}\) is predicted from the LSE of the HEA NP surface prior to adsorption.}
\end{figure*}

High-entropy alloys (HEAs), composed of five or more elemental species at concentrations ranging from 5 to 35 at\%, have emerged as versatile materials with promising applications in catalysis and as functional materials.\cite{cantor2004microstructural, yeh2004nanostructured, Pickering_2016, George_2019}. Their rich compositional diversity facilitates the way for the "cocktail effect", resulting in unexpected properties that often surpass those of traditional single-element systems.\cite{Pickering_2016, CAO2020250}. Recent advancements have highlighted their superior catalytic performance\cite{Wu_2020, wu2020platinum, Kusada_2020, wu2022noble}; however, the vast array of potential combinations of elements poses a significant challenge for experimental exploration.

To address this complexity, studies have focused on computational methods for efficient screening\cite{Batchelor_2019, pedersen2020high, Oh_2019, wu2022noble, huo2023high}. First-principles calculations, such as density functional theory (DFT)\cite{norskov2011density, mardirossian2017thirty, Shiota_2020, saidi2022emergence}, coupled cluster (CC) theory\cite{mcclain2017gaussian,gruber2018applying,zhang2019coupled} and many-body perturbation theory (MBPT)\cite{shavitt2009many,schimka2010accurate}, describe chemical reactions on solid surfaces with high accuracy. Volcano plots, derived from first-principles calculations, illustrate the optimal adsorption energy range for catalytic activity, balancing between excessively strong and weak interactions \cite{BLIGAARD2004206, medford2015sabatier, huo2023high}. However, the heterogeneous surfaces of HEAs complicate the molecular adsorption characteristics, making first-principles approaches computationally intensive..

To circumvent these limitations, neural network potentials (NNPs) based on the Behler--Parrinello framework~\cite{Behler_2007, Behler_2011, behler2015constructing,Smith_2017, Lee_2019} and graph neural networks~\cite{gilmer2017neural,Sch_tt_2018, schutt2018schnet, Xie_2018}, offer promising solutions. Universal NNPs can encompass extensive elemental diversity and achieve high computational efficiency while maintaining accuracy on par with that of DFT~\cite{Sch_tt_2018, schutt2018schnet, Chen_2019, zitnick2022spherical, Chen_2022, takamoto2022towards, batatia2022mace, deng_2023_chgnet, gasteiger2022gemnet, tran2023open,kovacs2023mace, batatia2023foundation, musaelian2023learning, liao2023equiformerv2, merchant2023scaling, shiota2024taming}. Recently, NNPs specializing in HEAs have emerged, made more lightweight through knowledge distillation~\cite{clausen2024adapting}. These advances have accelerated the prediction of catalytic properties; however, computational challenges remain.

In contrast, descriptor-based machine learning models offer scalability by predicting adsorption energies through generalized coordination numbers (GCNs), d-band centers, surface microstructural features, and local atomic environments, bypassing direct energy assessments~\cite{calle2015finding, dean2019unfolding, nanba2021element, lu2020neural, pedersen2020high, yang2022applications, Tamtaji_2023, roy2021machine, roy2024unravelling, Batchelor_2019}. These models have been proposed for predicting the adsorption energies of the remaining candidates by regressing the adsorption energies obtained from several first-principles calculations. Nonetheless, their applicability to HEAs is hampered by the complexity of alloy compositions and their dependence on extensive first-principles calculations.

To address these challenges, we propose a methodology for predicting the molecular adsorption energies on multi-element surfaces, such as HEAs, without direct adsorption-energy computations. Our approach focuses on a new descriptor that reflects the local reactivity of solid surfaces at atomic resolution. A key feature of the proposed method is its ability to predict the properties of multi-element systems using models constructed from data on single-element systems. We validated our method by comparing it with DFT for predicting the adsorption energies of CO on IrPdPtRhRu HEA NPs.

\section{\label{sec:level1} Methods}

In this section, we introduce a novel model that employs data on monometallic surfaces to predict molecular adsorption energies on multimetallic surfaces. In Section 2.1, we clarify the target systems and the problems addressed in this study. In Section 2.2, we introduce the LSE---a scalar descriptor that captures the atomic-level surface stability---which is the foundation of our model construction and prediction. Section 2.3 outlines the methodologies employed to develop and refine the prediction model using the newly introduced LSE descriptor and data derived from monometallic surfaces.

\subsection{\label{sec:level2} Target systems and problems}

Designing new HEAs comprising five elements selected from a pool of approximately 40 different elemental candidates results in approximately $6.58 \times 10^5$ possible combinations. Moreover, even for a given set of five elements, an exponentially large degree of freedom exists in the distribution of these elements in the actual alloy. Furthermore, the catalysts synthesized in practice and use are NPs, which differ from ideal surfaces in that they contain sites with different coordination numbers, such as corners, edges, and facets (Figure~\ref{fig:1}(b)). These diverse surface environments are the source of the cocktail effect, which contributes to the variability in catalytic properties. However, considering all these degrees of freedom when screening new catalyst candidates is not feasible. In this study, our objective was to identify the distribution of molecular adsorption energies, assuming that the elemental composition, size, and shape of the HEA NPs were predetermined.

We made the following assumptions regarding the structure and composition of HEA NPs: the structure is a truncated octahedron NP with 201 atoms in a face-centered cubic (fcc) arrangement (see~\ref{fig:1}(d)), the elemental composition ratio is as uniform as possible, and the atomic arrangement is randomly determined following a uniform distribution. We also assumed that the molecule occupied only a single top site. In HEA\(_{201}\), there are 122 sites. Even with these assumptions, the number of adsorption sites on HEA NPs is approximately $1.3 \times 10^{138}$ (Figure~\ref{fig:1}(a)). 

As a specific demonstration system, we measured the adsorption energies of CO molecules on IrPdPtRhRu NPs. In 2020, Wu et al. successfully synthesized IrPdPtRhRu NPs with nearly identical experimental composition ratios ~\cite{Wu_2020}. The investigation of the adsorption characteristics of CO molecules is useful for evaluating the catalytic properties of HEAs \cite{pedersen2020high, salinas2024toward}. As precursor systems for developing predictive models for HEAs, the monometallic NPs are of the truncated octahedron type corresponding to $\mathrm{M}_\mathrm{n}$ {n=38, 79, 116, 201}. For the on-top adsorption of the CO molecule on monometallic NPs $\mathrm{M}_{201}$, only the irreducible adsorption sites are calculated as shown in Figure~\ref{fig:1}(c). The adsorption energy $E_{\mathrm{ad}}$ is calculated as follows:
\begin{equation}
E_{\mathrm{ad}} = E_{\mathrm{tot}}^{\mathrm{CO/M_n}}-E_{\mathrm{tot}}^{\mathrm{CO}}-E_{\mathrm{tot}}^{\mathrm{M_n}},
\label{eq:4}
\end{equation}
where $E_{\mathrm{tot}}^{\mathrm{CO/M_n}}$, $E_{\mathrm{tot}}^{\mathrm{CO}}$, and $E_{\mathrm{tot}}^{\mathrm{M_n}}$ denote the total energies of $\mathrm{CO/M_n}$, $\mathrm{CO}$, and $\mathrm{M_n}$, respectively. The adsorption energies were computed using both the pretrained universal NNP M3GNet\cite{Chen_2022} and DFT at the Perdew--Burke--Ernzerhof (PBE) level. 

\subsection{\label{sec:level2}Local surface energy (LSE) descriptor}

\begin{figure}[h]
\centering
\includegraphics{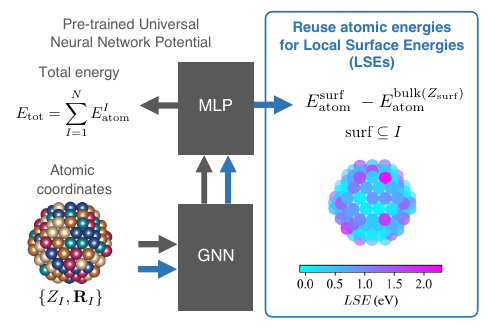}% Here is how to import EPS art
\caption{\label{fig:11} Framework for predicting Local Surface Energies (LSEs) using a pre-trained universal neural network potential (NNP). The gray arrow indicates the computational flow from the pre-trained universal NNP, and the blue arrow represents the flow that generates LSEs. Atomic coordinates \(\{Z_I, R_I\}\) are processed by a graph neural network (GNN) layer to extract atomic features. A multi-Layer perceptron (MLP) layer predicts the atomic energies \(E_{\text{atom}}^I\), which are reused to compute LSEs as the difference between surface atomic energies \(E_{\text{atom}}^{\text{surf}}\) and bulk atomic energies \(E_{\text{atom}}^{\text{bulk}}(Z_{\text{surf}})\). The resulting LSEs are visualized with a color map, highlighting local reactivity on the surface.}
\end{figure}

The LSE descriptor is defined as follows:
\begin{equation}
LSE \equiv E_{\mathrm{atom}}^{\mathrm{surf}}-E_{\mathrm{atom}}^{\mathrm{bulk}}, 
\label{eq:1}
\end{equation}
where $E_{\mathrm{atom}}^{\mathrm{surf}}$ and $E_{\mathrm{atom}}^{\mathrm{bulk}}$ denote the atomic energies in the surface and bulk environments, respectively. The LSE represents the energy loss caused by a single atom in a single-element (or unary) bulk environment when exposed to a single- or multi-element surface. This definition enables the quantification of the surface stability, even for surfaces in complex environments and multicomponent systems. The atomic energies in Equation (\ref{eq:1}) can be evaluated via energy density analysis (EDA) from first-principles calculations, such as DFT\cite{Nakai_2002, kikuchi2009one, Yu_2011, Huang_2019, Eriksen_2020}, which was introduced by Nakai in 2002 \cite{Nakai_2002}. EDA is accurate because it is based on first-principles calculations; however, it is not suitable for exhaustive calculations, such as those in the present study, because of its high computational cost. To reduce the computational cost, all the LSE values in this study were evaluated using a machine learning interatomic potential (MLIP), specifically the universal NNP M3GNet. In our previous study, we demonstrated that the intermediate information from M3GNet can efficiently and accurately represent the local environments of atoms in molecules.\cite{shiota2024universal} In the Behler--Parrinello NNP framework, the total energy $E_{\mathrm{tot}}$ of a system comprising $N$ atoms is calculated as the sum of the energies of the atoms.
\begin{equation}
E_{\mathrm{tot}}=\sum_{I=1}^N E^I_{\mathrm{atom}}
\label{eq:2}
\end{equation}
Yoo et al. demonstrated the ability to map the atomic energies obtained by NNPs onto NP and surface systems \cite{Yoo_2019}. Deringer et al. utilized the Gaussian approximation potential model to compute the atomic energies, which were then used to explore the configurational space and investigate the nature of defects in crystals \cite{el2022exploring, morrow2024understanding}. MLIPs such as NNPs enable efficient evaluation of LSEs because the atomic energies of all adsorption sites in one system can be obtained in a single calculation. Figure \ref{fig:11} illustrates the workflow for evaluating the Local Surface Energy (LSE) using a pre-trained universal NNP.

\subsection{\label{sec:level2}Prediction model based on monometallic data}

We introduce a predictive model for the adsorption energy of a molecule on a multi-element surface. This model is defined as a regression between the molecular adsorption energy on monometallic surfaces for each constituent element $\mathrm{M}$ of the multimetallic system and the LSE of the adsorption site prior to molecular adsorption. Figure \ref{fig:1}(c) illustrates the workflow for constructing the predictive model. As the simplest model, we adopted the least-squares linear regression model expressed as follows:
\begin{equation}
E_{\mathrm{ad}}^\mathrm{M}(\mathrm{Predict.}) = \alpha_{\mathrm{M}} \times LSE +\beta_{\mathrm{M}}.
\label{eq:3}
\end{equation}
Here, $\alpha_{\mathrm{M}}$ and $\beta_{\mathrm{M}}$ denote the regression coefficients and constants, respectively, for each element $\mathrm{M}$. Simple regression makes the model explainable. $\alpha_{\mathrm{M}}$ represents the magnitude of the adsorption energy response to a change in LSE. $\beta_{\mathrm{M}}$ represents the adsorption energy of a molecule when the LSE is 0, that is, when the surface atom has the same energy as that in the bulk environment. For prediction, the adsorption energy is estimated by substituting the LSE values of the multi-element alloy surface prior to molecular adsorption into Equation (\ref{eq:3}). Figure \ref{fig:1}(d) illustrates the workflow for predicting CO adsorption energies on HEA NPs using the proposed model. Here, the prediction model using adsorption energies from the NNP for regression is referred to as the LSE-based prediction model, while the model using adsorption energies from DFT is referred to as the Improved-LSE (I-LSE) prediction model. Details of the model construction can be found in the Computational details section. 

\begin{figure}[h]
\includegraphics{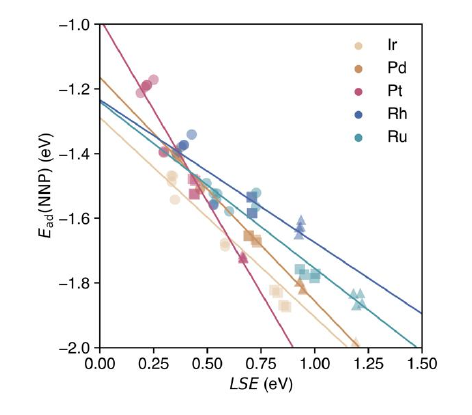}% Here is how to import EPS art
\caption{\label{fig:2}Adsorption energies of CO calculated using the NNP for each on-top adsorption site of monometallic NPs $\mathrm{M}_\mathrm{n}$ with respect to the LSE. Solid lines represent the linear regressions of the adsorption energies of a CO molecule at the on-top sites of each monometallic NP according to Equation (\ref{eq:3}). Circles, squares, and triangles at each datapoint represent the facet, edge, and corner CO adsorption sites, respectively.}
\end{figure}

\begin{figure}[h]
\includegraphics{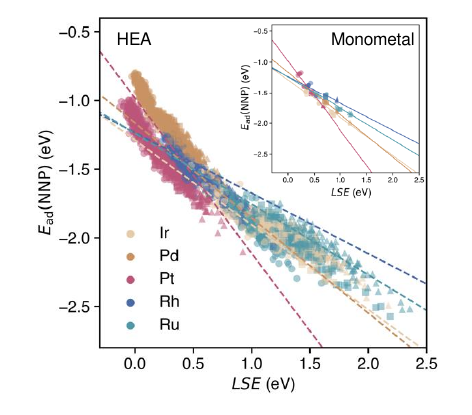}% Here is how to import EPS art
\caption{\label{fig:3}Adsorption energies of CO calculated using the NNP for each on-top adsorption site of 20 HEA$_{201}$ NPs with respect to the LSE. The inset presents a comparison of the distribution of adsorption energies for CO between the HEA NPs and monometallic NPs, with the x- and y-axes rescaled from Figure~\ref{fig:2} for consistency. Dashed lines represent the linear regressions of the adsorption energies of a CO molecule at the on-top sites of each monometallic NP based on Equation (\ref{eq:3}). Circles, squares, and triangles at each datapoint represent the facet, edge, and corner CO adsorption sites, respectively.}
\end{figure}

\begin{figure*}[h]
\includegraphics{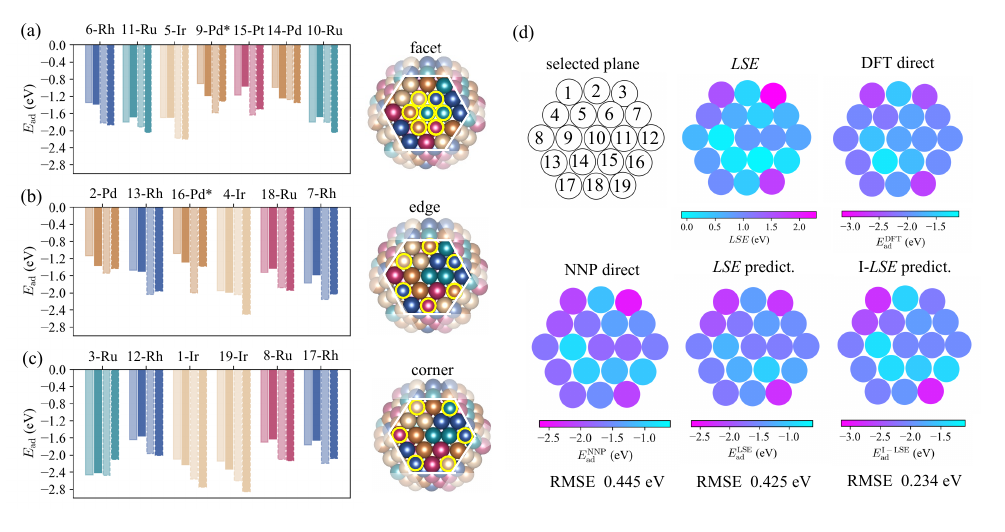}% Here is how to import EPS art
\caption{\label{fig:4}Bar graphs in (a), (b), and (c) show the evaluated adsorption energies of CO on the (111) plane of a randomly selected $\text{HEA}_{201}$ at each on-top site of the facet, edge, and corner, respectively. The atoms of the selected plane are numbered as shown in (d). For each adsorption site, from left to right, the bar graph represents the adsorption energy obtained via direct evaluation using NNP (NNP direct), LSE-based prediction (LSE predict.), DFT, and LSE-based prediction parameterized by DFT data (I-LSE predict.). (d) Color mapping of the LSE values of the atoms on the selected plane and corresponding adsorption energies shown in (a), (b), and (c), along with the RMSE values relative to the DFT results. The asterisk indicates that the results of the structural optimization converge on the bridge site rather than the on-top site.}
\end{figure*}

\section{\label{sec:level1}Results}

In this section, we present the results for the prediction of the adsorption energy of HEA NPs using the proposed methods based on the LSE descriptor, focusing on the computational efficiency and precision. In Section 3.1, we demonstrate a strong correlation between the adsorption energy and the LSE descriptor, confirming that the LSE is a reliable descriptor of the adsorption energy between different sites in the NP. In Section 3.2, we verify the accuracy of LSE-based predictions by comparing them with DFT calculations, thereby proving the robustness of the prediction model. In Section 3.3, we analyze how the diverse surface structures and elemental compositions of 1000 different HEA NPs, comprising 122000 environments, lead to a wide range of adsorption-energy distributions. Finally, in Section 3.4, we compare the computational efficiency of our LSE-based adsorption energy predictions with that of direct adsorption-energy predictions using the traditional NNP and DFT methods.

\subsection{\label{sec:level2}Correlation Between Adsorption Energy and LSE}

\begin{figure*}[h]
\includegraphics{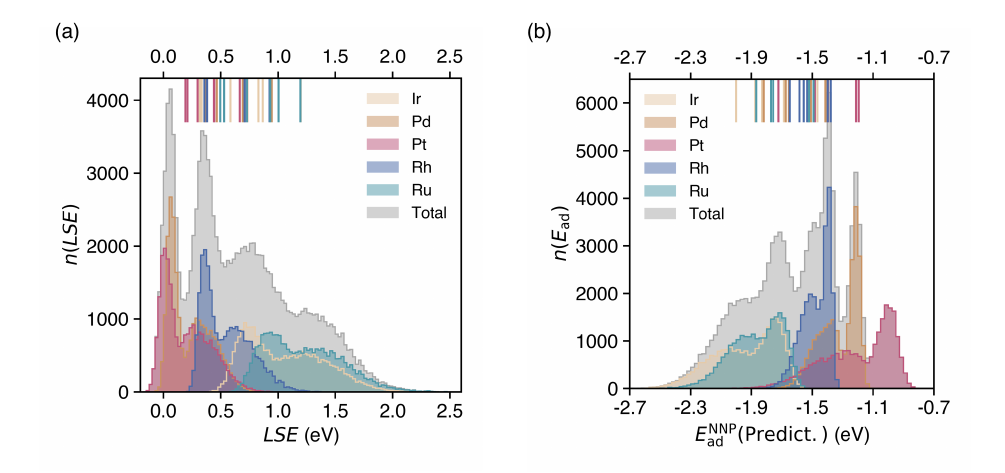}% Here is how to import EPS art
\caption{\label{fig:5}(a)Distributions of 122,000 LSE values of all the topmost layer atoms of the 1000 structure-optimized $\mathrm{HEA}_{201}$ NPs for each element and their sum, respectively. (b) Distributions of adsorption energies of CO for all the on-top adsorption sites predicted using the LSEs in (a) and Equation (\ref{eq:3}). Solid lines represent the values for monometallic NP M\(_{201}\) (M=Ir, Pd, Pt, Rh, and Ru).}
\end{figure*}

The correlation between the LSE and the CO adsorption energy $E_{\mathrm{ad}}(\mathrm{NNP})$ of the monometallic NPs obtained using the NNP is shown in Figure \ref{fig:2}. The solid lines represent linear regressions of the adsorption energies of a CO molecule at the top sites of each monometallic system. For all the metal elements, the relationship between the adsorption energy $E_{\mathrm{ad}}(\mathrm{NNP})$ of the CO molecule at the top site and the LSE is linear. The LSE values of all the adsorption sites for all elements range from approximately 0.2 to 1.2 eV. In other words, in all cases, the atomic energies are more unstable in the surface environment than in the bulk environment, which is reasonable given the lower coordination number in the surface environment. The adsorption energy $E_{\mathrm{ad}}(\mathrm{NNP})$ ranges from approximately -1.2 to -2.0 eV. Next, we examined the adsorption sites on the NPs. For all elements, the adsorption energy decreased in the following order: facets, edges, and corners. In contrast, the LSE values increased in the order of facets, edges, and corners. The LSE values and adsorption energies were concentrated at the facets, edges, and corners, and energy gaps existed between each group of adsorption sites. The $\mathrm{RMSE}$ was 0.035 eV, revealing a strong correlation between the LSE and the adsorption energy (see Figure~\ref{fig:6} in the Details of the LSE-based regression models in Appendix). 

Figure~\ref{fig:3} presents the adsorption energies calculated directly using the NNP and their predicted values (dashed lines). The RMSE was 0.150 eV, which exceeded that of the unitary system. However, a strong correlation was observed between the LSE and adsorption energy in the HEA, indicating that the adsorption energy in a multicomponent environment can be effectively predicted (Figures~\ref{fig:8}(a) and (b) in the Appendix). This finding suggests that the LSE can efficiently and accurately predict the adsorption energies not only for unitary systems but also for complex systems such as HEAs. Notably, when we used the adsorption-energy range of -2.0 to -1.2 eV in the unitary system as the interpolation region for ${E_\mathrm{ad}(\mathrm{Predict.})}$, the predictions were more reliable than those for other ranges. In the extrapolation region of ${E_\mathrm{ad}(\mathrm{Predict.})}$, the difference from ${E_\mathrm{ad}(\mathrm{NNP})}$ increased, and a maximum shift of approximately 0.5 eV was observed. 

Next, we explored the trends for each elemental species. Figures~\ref{fig:3} and ~\ref{fig:8}(a) and (b) in the Appendix show that for all elemental species, the adsorption energies are nonlinearly estimated toward the unstable adsorption energy side. Additionally, the nonlinear region was dominated by adsorption at the facet sites. Pd atoms in HEA nanoparticles clearly show this non-linear behavior; unlike Pt atoms, they show a reversed pattern of adsorption energies. Compared to sites with low coordination numbers, such as corner and edge sites, the atoms on the facets were coordinated with eight or nine atoms. This increased coordination number renders them more sensitive to the surrounding environment than unitary systems. Consequently, the complex environment of HEAs may introduce unexpected nonlinearity into predictions. To show that this nonlinear trend can be captured by training the adsorption energies on HEA NPs, we applied kernel ridge regression (KRR) and quantum circuit learning (QCL) regression \cite{Mitarai_2018} to construct an adsorption energy prediction model for each elemental species at the HEA NP adsorption sites. Figures~\ref{fig:8}(c) and (d) in the Appendix show the correlation plots between the predicted and actual adsorption energies of CO on 14 HEA NP patterns generated by the regression model using KRR and QCL regression, respectively. The RMSEs of the adsorption energy predictions for all the adsorption sites provided by the KRR and QCL regression models were 0.0580 and 0.0579 eV, respectively, indicating comparable nonlinear regression performance. Compared with the uncorrected predictions based on the LSE, the RMSE values were reduced by a factor of approximately three.

\subsection{\label{sec:level2}DFT verification of the LSE-based predictions}

The accuracy of the adsorption energy predictions was verified using an NNP and the LSE with DFT calculations while seeking to increase the prediction accuracy. In our computing environment, the computation time for evaluating the adsorption energy via DFT was approximately $10^3$ times that of the NNP, as indicated by Table~\ref{tab:1}. Therefore, we randomly selected one of the 20 structures of $\text{HEA}_{201}$ discussed above and evaluated the adsorption energies for 19 sites on the (111) plane, as shown on the right side of Figure~\ref{fig:4}. The selected $\text{HEA}_{201}$ was $\text{Ir}_{40}\text{Pd}_{40}\text{Pt}_{41}\text{Rh}_{40}\text{Ru}_{40}$. First, the prediction accuracy of the adsorption energies between the prediction based on the LSE and direct NNP calculation was compared with that of the DFT calculations. The RMSE of the adsorption energy for all sites obtained via the direct NNP calculation was 0.445~eV. The RMSE value of the prediction based on the LSE was 0.425~eV, corresponding to a slightly higher accuracy (0.020~eV) compared with the direct evaluation via the NNP. Thus, the accuracy of the prediction model using the LSE was close to that of the NNP. Although the RMSE of 0.445~eV for direct NNP calculations may appear large, the M3GNet NNP systematically overestimates CO adsorption energies, with a mean error (ME) of 0.388~eV relative to DFT as shown in Figure~\ref{fig:4}. A similar overestimation trend is observed for monometallic NPs, where the RMSE and ME against DFT are 0.439~eV and 0.379~eV, respectively. As a result, the LSE-based prediction also displays a systematic overestimation of the CO adsorption energy in comparison with DFT. Notably, the adsorption energy range for the 19 HEA sites spans from 1.329~eV to 1.567~eV (by DFT and NNP, respectively), indicating a broad distribution of possible adsorption energies. Given this wide range, a constant shift between NNP and DFT remains comparatively tolerable for high-throughput screening of HEA catalysts.

While the M3GNet NNP has difficulty quantitatively describing CO-adsorbed states, it effectively captures the atomic-level stability of the NPs. This is supported by the small RMSD (0.09~\AA) between the DFT-relaxed and NNP-relaxed HEA NP structures, as well as the LSE RMSE of 0.026~eV for the 19 target sites, indicating that M3GNet—trained on diverse crystal environments—can reliably describe complex HEA NP configurations despite their absence from its original training set. Consequently, by combining DFT-derived adsorption energies for monometallic NPs with M3GNet-derived LSE (i.e., the I-LSE method), the prediction error is halved to an RMSE of 0.234~eV. This improvement arises because the LSEs are calculated from the NPs before adsorption, thereby avoiding the systematic overestimation of adsorption energy found in direct NNP calculations, while retaining the efficiency of an LSE-based framework.

\subsection{\label{sec:level2}Distribution of the Adsorption Energy on HEA NPs}

Figure \ref{fig:5}(a) shows the 122000 LSE values of 1000 HEA NPs for each element and their total distribution. This distribution becomes smoother and converges as the number of NP patterns increases. The sum of all the elemental distributions shown in gray in Figure~\ref{fig:5}(a) can be considered as an indicator of the reactivity of the entire HEA NP surface of the given elements. In the monometallic system, the LSE ranged from approximately 0.2 to 1.2 eV. However, in the quintic HEA environment, these values underwent significant changes and ranged from approximately -0.1 to 2.4 eV. The distribution of each element exhibited two prominent peaks. For Pt and Pd, the LSE exhibited a major peak near 0 eV (or a slightly lower energy), indicating improved stability compared with that in the monometallic environment. Second, the smaller peak remained nearly unchanged for Pd, whereas the LSE range expanded by approximately 0.3 eV for Pt. For Ir and Ru, the LSE values shifted toward higher energies compared with those in the monometallic systems. For Rh, a slight increase in the LSE range was observed, lying between those of the stable Pt and Pd groups and the less stable Ir and Ru groups.

Figure~\ref{fig:5}(b) presents the predicted adsorption energies of CO on all the on-top sites of 1000 $\mathrm{HEA_{201}}$ NPs obtained using Equation (\ref{eq:3}) with LSE values. The range of the on-top adsorption energy $E_\mathrm{ad}$ for CO in the monometallic system (Figure \ref{fig:2}) expanded by approximately 0.8 eV, from approximately -2.0 to -1.2 eV to -2.5 to -0.9 eV. This serves as an example of how adsorption characteristics diversify in a quintic HEA environment. The distribution of the adsorption energies for each element also exhibited two prominent peaks. Next, we examined the differences between the elemental species. For Pd and Pt, the LSE values were very close, but the range of adsorption energies was broader on the high-energy side by 0.2 eV for Pt. Similar trends were observed for Ir and Ru on the low-energy side. In the case of Rh, a slight extension in the range of both the high- and low-energy sides was observed compared with that of the monometallic system. In particular, the adsorption energy was concentrated in three adsorption site groups (corners, edges, and facets) in the monometallic NP environment, but these groups exhibited a broader range of values in the HEA NP environment. This representation as a distribution can help characterize and visualize the potential cocktail effect for efficient screening of novel HEAs across the periodic table.

\subsection{Universality of the LSE-based method}

\begin{figure*}[]
\includegraphics{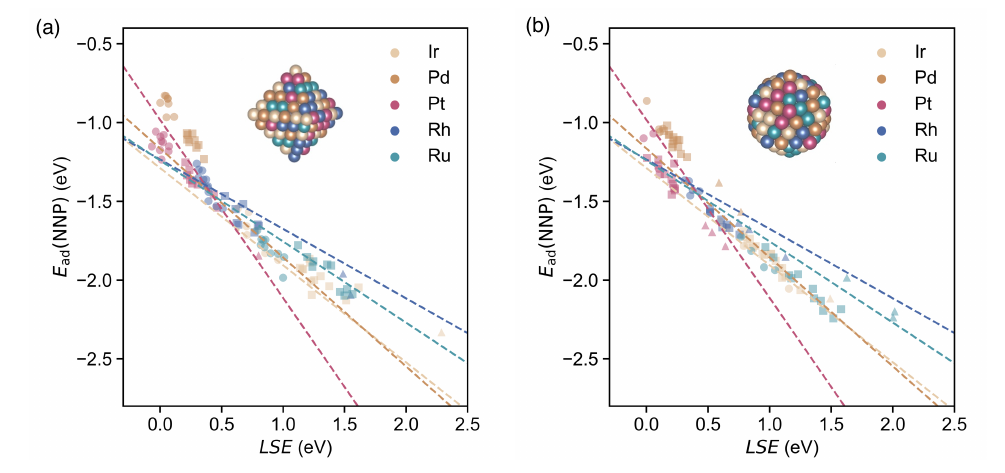}% Here is how to import EPS art
\caption{\label{fig:10}Universality of the LSE-based prediction model for CO adsorption energy across different HEA nanoparticle (NP) shapes. (a) Predicted adsorption energies from the NNP for all ontop sites on a regular octahedron-shaped HEA$_{146}$, plotted as a function of LSE. (b) Corresponding predictions for an icosahedron-shaped HEA$_{147}$. The dashed lines represent the LSE-based linear regression model constructed for monometallic truncated octahedron NPs.}
\end{figure*}

In this section, we discuss the universality of the LSE-based prediction model constructed in this study. As described in Section 3.1, due to the vastness of the chemical space for HEA NPs, we restricted both the model construction and the target NP shape to truncated octahedra. However, the LSE-based model can be universally applied to any HEA NP shape, provided the structure is given. This is because the universal NNP is capable of evaluating atomic energies for the given structure and thus computing the LSEs. The resulting LSEs can then be fed into the LSE-based model—originally constructed using monometallic data—to predict adsorption energies for the given HEA NP.

To verify the shape-independence of LSE-based prediction models, we evaluated its performance on two test sets: HEA$_{146}$ in the form of a regular octahedron and HEA$_{147}$ in the form of an icosahedron, as shown in the respective subsets in Figures \ref{fig:10}(a) and (b). These shapes were selected because they are among the most prevalent geometries for fcc NPs. Figures \ref{fig:10}(a) and (b) present the predicted CO adsorption energies from the NNP for all possible ontop sites on the regular octahedron and icosahedron, respectively. The corresponding LSE-based predictions are represented by the dashed lines as a function of LSE. For both NP shapes, we found clear correlations with the predictions obtained from the LSE-based model. The RMSE were 0.156 eV for the regular octahedron and 0.142 eV for the icosahedron, comparable to the RMSE of 0.150 eV achieved for truncated octahedra. These results demonstrate that our predictive framework exhibits universality with respect to NP shape within this scope. 

It should be noted that while regular octahedra and icosahedra showcase the applicability of the LSE-based method for commonly explored fcc morphologies, many other potential NP shapes and surface reconstructions remain unexplored in this study. We focus here on the representative geometries to illustrate the applicability of the method, recognizing that a comprehensive proof of “universality” would require testing across more diverse crystal structures and potential defect sites of the HEA NPs. These aspects, along with extensions to other chemical compositions and surface features, are promising avenues for future research.

\subsection{Comparison with GCN-based prediction model}

\begin{figure*}[]
\includegraphics{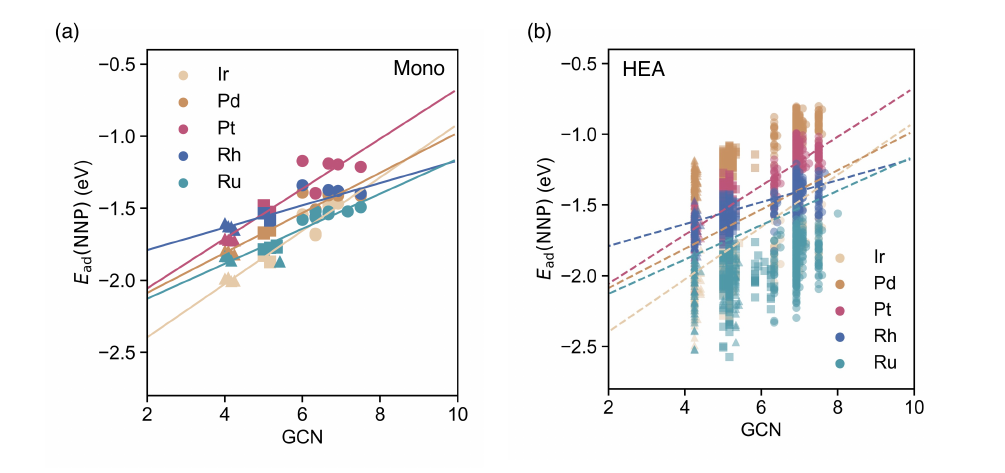}% Here is how to import EPS art
\caption{\label{fig:9}(a) Adsorption energies of CO calculated using the NNP for each on-top adsorption site of monometallic NPs $\mathrm{M}_\mathrm{n}$ with respect to the GCN values. Solid lines represent the linear regressions of the adsorption energies of a CO molecule at the on-top sites of each monometallic NP and the GCN values. (b) Adsorption energies of CO calculated using the NNP for each on-top adsorption site of 20 HEA$_{201}$ NPs with respect to the GCN values. The dashed lines represent the same linear regressions as in (a). Circles, squares, and triangles at each datapoint represent the facet, edge, and corner CO adsorption sites, respectively.}
\end{figure*}

In this section, we employ a GCN descriptor—a scalar descriptor similar to LSE—to construct a prediction model based on monometallic data and compare its predictive accuracy. The GCN descriptor quantifies the environment of the adsorption site by counting the coordination number of the nearest-neighbor atoms, and it is defined in Ref.~\cite{calle2015finding} Following the same approach used for the LSE-based predictive model of Equation (\ref{eq:3}), we examined the correlation between CO adsorption energies on monometallic surfaces and GCN values. Linear regression models were then constructed for each element type. As shown in Figure \ref{fig:9}(a), similar to LSE, a linear model accurately represents the relationship, achieving a precision of approximately 0.06 eV. The parameters of the linear regression models are summarized in Table 4 in Appendix section.

Subsequently, we applied these linear models to predict the adsorption energies of the 20 HEA NPs presented in the Results section to evaluate the prediction accuracy. As depicted in Figure \ref{fig:9}(b), although the GCN-based predictions correlate with the directly computed adsorption energies, the discrete nature of GCN leads to variations of approximately from 1 to 1.5 eV even for identical GCN values. The resulting RMSE was 0.278 eV, approximately twice as large as that obtained using the LSE-based predictions, indicating that GCN alone is insufficient for capturing the environmental changes in multicomponent systems. This trend is consistent with findings reported by Namba et al.~\cite{nanba2021element} Nonetheless, future improvements may be possible by combining GCN with other descriptors, such as LSE, to enhance the predictive capabilities of monometallic data-based adsorption energy prediction models.

\subsection{\label{app:2}Computational Efficiency}

\footnotetext{\footnote[1]~NNP calculations with M3GNet were performed using an AMD EPYC 7532 32-core processor with 64 CPUs.}
\footnotetext{\footnote[2]~DFT calculations with VASP were performed using 10 Intel(R) Xeon(R) Platinum 9242 CPUs each with 96 CPUs in an MPI parallel configuration, for 960 CPUs.}

\begin{table}[b]
\caption{\label{tab:1}%
Computational costs for 19 geometry optimizations of CO adsorption on the (111) plane on $\mathrm{Ir_{40}Pd_{40}Pt_{41}Rh_{40}Ru_{40}}$ HEA NP via NNP and DFT.}
\begin{tabular}{cccc}
&total time[sec.]& total step[step] &sec./step\\
%\mbox{Three}&\mbox{Four}&\mbox{Five}\\
\hline
NNP\footnote[1]{NNP calculations with M3GNET were performed using an AMD EPYC 7532 32-core processor with 64 CPUs.}&2314&10480&0.22\\
DFT\footnote[2]{DFT calculations with VASP were performed using 10 Intel(R) Xeon(R) Platinum 9242 CPUs each with 96 CPUs in an MPI parallel configuration, for 960 CPUs.}&925518&5008&184.81\\
%NNP/DFT\footnote{NNP/DFT represents the ratio of values obtained using the NNP to that calculated using DFT.}&$3\times 10^{-3}$&2&$1\times 10^{-3}$\\
\hline
\end{tabular}
\end{table}

Finally, we compared the computational performance of our method with that of conventional approaches for obtaining adsorption energy distributions. The time required to predict the adsorption energies of CO at 122000 sites across 1000 patterns of $\mathrm{HEA}_{201}$ using LSE was compared with the time required for direct calculations using the NNP and DFT. The use of the LSE eliminates the need for optimization of the adsorption structure for CO molecules. The computation times for the NNP and DFT are presented in Table~\ref{tab:1}. The computation time per structural optimization step with the NNP was approximately 1000 times shorter than that with DFT. However, as mentioned in Section 3.4, performing direct structural optimizations for 122000 adsorption sites requires approximately 171 days using the NNP and 188 years using DFT. These lengthy timescales were avoided by using the proposed method. The most time-consuming process in our approach is obtaining the LSE values through NNP structural optimization of 1000 HEA NPs, which was completed in just 1.4 days. The significant reduction in the execution time for adsorption energy predictions achievable with this method is expected to facilitate the screening of catalytic properties for HEAs with exponentially large combinations using the entire periodic table as candidates.

\section{\label{sec:level1}Discussion}

We utilized the atomic energy obtained from the NNPs and introduced a metric called the LSE, which represents the surface energy per atom. Using the LSE, we determined the adsorption energy of CO molecules on the top sites of IrPdPtRhRu $\mathrm{HEA_{201}}$ NPs with a large number of atomic combinations as a distribution (Figure~\ref{fig:5}(b)). This approach offers a novel means of analyzing the atomic energies to evaluate various adsorption energies in multicomponent systems such as HEAs. The adsorption-energy distribution obtained via LSE prediction can help efficiently and effectively  characterize and visualize unexpected cocktail effects induced by the vast chemical space of HEA NPs.

Notably, our calculations solely considered the adsorption energies on monometallic NPs and isolated multicomponent NPs, without the need for a direct evaluation of the CO molecular adsorption energy on the HEA. This approach enables the evaluation of adsorption energies approximately $10^5$ times faster than direct DFT calculations, facilitating the visualization of the cocktail effect. Regarding accuracy, a comparison with DFT calculations revealed that the predictions based on the LSE were nearly an order of magnitude larger than the chemical accuracy, although they were not quantitatively accurate.  Nonetheless, the relative energies exhibited a similar trend, indicating that qualitative comparisons that consider the influence of the surrounding environment on each element are feasible. Consequently, this method can be employed as a screening tool prior to applying DFT calculations or high-level quantum chemical methods such as CC theory.

The LSE-predicted adsorption energies were highly accurate, with an RMSE of 0.150 eV relative to the correct values, despite their low cost compared with direct evaluations. However, nonlinearity relative to the correct values was observed, indicating nonlinear behavior with respect to the LSE. We introduced a naïve method to capture this nonlinearity: nonlinear regression between the directly evaluated CO adsorption energy and the LSE. We modeled nonlinear regression with KRR and the classical quantum hybrid algorithm QCL regression. Learning the adsorption energies for only 732 sites for NPs in six patterns of structures improved the resulting LSE predictions thrice for the remaining 14 patterns tested for 1708 sites. The constraint that the norm of the parameters in QCL regression must equal 1 is expected to act as regularization. The results of the KRR and QCL regression models indicated similar regularization capabilities. Although our adsorption-energy correction model does not inherently require quantum computing, its utility may be extended to a wider chemical space and the construction of models encompassing the entire periodic table. Furthermore, in this study, the adsorption energies of quintet systems were studied; it may be possible to systematically increase the accuracy by adding binary and ternary adsorption energies to the model.

Finally, we discuss potential applications of the proposed method for designing chemical reactions for catalyst and device development. Previous studies examined the role of atomic energy in increasing the accuracy and efficiency of NNPs and validated atomic-energy mapping results \cite{Yoo_2019, Jeong_2020, Watanabe_2020, Doan_2023, chong2023robustness}. Recently, studies have attempted to gain chemical insight from atomic energies \cite{Kjeldal_2023}. However, to the best of our knowledge, the proposed LSE prediction method is the first to demonstrate that atomic energy can serve as a descriptor for the efficient evaluation of catalytic properties. Moreover, in contrast to atomic energies, which are absolute quantities, LSEs are relative quantities and are thus expected to be less sensitive to differences in computational methods, such as the treatment of basis functions and inner-shell electrons. The proposed approach enhances chemical reaction design—a crucial component of machine learning-based material design—which has garnered significant attention in the scientific community over the past few years due to its promise to accelerate the discovery and development of useful materials.\cite{Meuwly_2021, Wen_2023}.

\section{\label{sec:level1}Conclusion}
We developed a computational methodology for predicting the molecular adsorption energies on HEAs using the LSE descriptor derived from the NNPs calculated atomic energies. This method addresses the challenge of evaluating the vast chemical space of HEAs due to their compositional diversity and the computational expense associated with direct DFT and NNP calculations. The LSE descriptor efficiently captures the local reactivity of surface atoms, enabling rapid and accurate prediction of adsorption energies across a wide range of HEA configurations.

Our approach significantly accelerates the computational process, reducing the computation time from hundreds of years with DFT and hundreds of days with the NNP to only a few hours, which makes it a practical tool for materials discovery and catalyst design. The adoption of nonlinear regression techniques combined with advanced machine learning models, such as KRR and QCL regression, has increased the accuracy of adsorption-energy predictions, even in the face of the nonlinearity inherent in multicomponent systems.

Building upon this work, future research can extend the application of the LSE descriptor to other molecular species and reaction systems, validating and enhancing its predictive accuracy across a broader spectrum of catalytic processes. Integrating the LSE descriptor with advanced machine learning algorithms could facilitate large-scale screening of HEA compositions, accelerating the discovery of optimal catalysts for specific reactions. Moreover, combining the LSE descriptor with additional descriptors, such as surface microstructural features or GCN, offers the potential to further refine predictive accuracy by accounting for local atomic environments in more detail.\cite{calle2015finding, dean2019unfolding, nanba2021element, lu2020neural, pedersen2020high, yang2022applications, Tamtaji_2023, roy2021machine, roy2024unravelling, Batchelor_2019}

In this study, we assumed that the HEA NPs have random elemental configurations following a uniform distribution. However, short-range order is also important in high-entropy alloys\cite{singh2015atomic,sun2022effect,sheriff2024quantifying}. By combining our method with techniques for quantifying short-range order using E(3)-equivariant graph neural networks \cite{sheriff2024chemical}, we can potentially obtain more accurate distributions of reactivity. This integration would allow us to capture the effects of atomic arrangements more precisely, leading to improved predictions of catalytic properties.

Moreover, this approach holds the potential to consider more realistic environmental conditions, including the behavior of catalysts in solution or under operational settings. Incorporating factors such as solvent effects, temperature, and pressure into the screening process would enhance the relevance and applicability of the predictions, leading to more effective and practical catalyst designs. Furthermore, applying the LSE framework to other material systems, such as high-entropy nitrides, oxides, and carbides, could further expand its impact on materials design.

In conclusion, this research not only paves the way for rapid and accurate computational screening of catalytic materials but also sets the stage for developing computational tools capable of handling the complexities of modern materials science—particularly in the realm of high-entropy materials.

\section{\label{sec:level1}Computational Details}

\subsection{\label{sec:level2}Modeling}

We modeled CO adsorption on the NPs, i.e., CO/$\mathrm{M}_{\mathrm{n}}$, using an atomic simulation environment (ASE) \cite{ISI:000175131400009, ase-paper}. The initial lattice constants (LCs) of $\mathrm{M}_{\mathrm{n}}$ were determined via bulk calculations. For $\mathrm{HEA}_{\mathrm{201}}$, the largest bulk was used as the initial LC. A 15-Å vacuum region was inserted in all supercells to minimize cell-to-cell interactions. The initial structures of $\mathrm{CO/M}_{\mathrm{n}}$ and $\mathrm{CO/HEA}_{\mathrm{201}}$ were derived by placing CO on the on-top sites of the optimized structures of $\mathrm{M}_{\mathrm{n}}$ and $\mathrm{HEA}_{\mathrm{201}}$, respectively, with the distance between the C atom and the adsorption-site metal atom M set as 2 Å. 

\subsection{\label{sec:level2}NNP calculations}

The NNP used was M3GNet, a universal NNP proposed by Chen and Ong~\cite{Chen_2022}. This M3GNet NNP was trained on approximately 180,000 crystal environments at the PBE or PBE+U levels of theory from the Materials Project~\cite{10.1063/1.4812323}, covering 89 elements. The crystal structures of the bulk $\mathrm{M=Ir, Pd, Pt, Rh, Ru}$ were assumed to be fcc. Although $\mathrm{Ru}$ exhibits a hexagonal close-packed form at room temperature, $\mathrm{Ru}$ with an fcc structure can be created using NPs \cite{Kusada_2013}. The atomic energy of the bulk fcc metal M was determined using the energy corresponding to the minimum value obtained when varying the LC of the material at intervals of approximately 0.01 $\mathrm{\AA}$. The resulting atomic-energy values and corresponding LCs are presented in Table~\ref{tab:2}. The Broyden--Fletcher--Goldfarb--Shanno (BFGS) algorithm was used for structural optimizations, with a maximum step of 0.005 Å. In the CO/M and CO/HEA structural optimization calculations, the M and HEA structures were fixed as stable isolated systems. Only the CO and adsorption-site atoms were relaxed, with the constraint that the adsorbed molecules occupy the top site. Structural optimizations were performed until the force acting on each atom was 0.001 eV/Å.

\subsection{\label{sec:level2}DFT calculations}
The DFT calculations were performed using the Vienna Ab initio Simulation Package (VASP), version 5.4.4 \cite{Kresse_1993, PhysRevB.54.11169, KRESSE199615}. We employed the PBE generalized gradient approximation functional as the exchange-correlation functional\cite{Perdew_1996}. The core electrons were treated using the projector-augmented wave method \cite{Bl_chl_1994, Kresse_1999}. Electronic structures were optimized using the blocked Davidson iteration scheme within a spin-restricted approximation. The cutoff energy for the plane-wave functions was set as 400 eV. Atomic coordinates were optimized using a conjugate-gradient algorithm with a convergence threshold of 0.01 eV/Å. 

\subsection{\label{sec:level2}GCN evaluations}
The GCN descriptor is evaluated according to the following equation, which is identical to the one defined in Ref.~\cite{calle2015finding}

\begin{equation}
\text{GCN}(i) = \sum_{j=1}^{n_i} \frac{cn(j)}{cn_{\max}}.
\label{eq:5}
\end{equation}
Here, \(i\) represents the index of the surface atom of the NPs. \(n_i\) is the coordination number of the \(i\)-th atom, and \(cn_{\max}\) is the coordination number of the \(i\)-th atom in its bulk environment, which is 12 for FCC metals considered in this study. \(cn(j)\) represents the coordination number of each of the \(n_i\) atoms coordinated to the \(i\)-th atom. In this study, GCN was evaluated for structures optimized using the universal NNP M3GNet, following the same procedure as for the evaluation of the LSE.

\subsection{\label{sec:level2}Nonlinear regressions}

To capture the nonlinearity between LSE and CO adsorption energy, we trained a nonlinear regression model using all possible ontop adsorption sites (732 in total) from six randomly selected HEA nanoparticles out of the 20 used for validating the linear LSE-based model. For validation, all possible ontop adsorption sites on the remaining 14 HEA nanoparticles were utilized. KRRs were executed using scikit-learn version 1.2.2 \cite{scikit-learn}. A Gaussian kernel was selected as the kernel function of the KRR. The hyperparameters for each model were optimized over 100 iterations using a randomized search. QCL regressions were implemented using scikit-qulacs version 0.5.0\cite{qulacs_osaka}. Qulacs version 0.5.6 was used as the quantum-circuit simulator\cite{Suzuki_2021}. The number of qubits in the QCL regression model was 4. The number of iterations of the parameterized variational quantum circuit of the model corresponding to the weights of the neural network was 6. The timestep for the time evolution operator in parameterized variational quantum circuits was set as 0.5. The BFGS algorithm was used to update the parameters of the QCL model.

\begin{table}[]
\caption{\label{tab:2}Atomic energies $E_{\mathrm{atom}}^{\mathrm{bulk}}$ and lattice constants (LC) of face-centered cubic (fcc) bulk metals M (M = Ir, Pd, Pt, Rh, Ru), obtained from M3GNet calculations and used for computing the local Surface Energy (LSE). The atomic energies are the minimum values determined by varying the lattice constant in intervals of approximately 0.01 Å.}
\begin{tabular}{cccccc}
fcc bulk M&Ir&Pd&Pt&Rh&Ru\\
\hline
$E_{\mathrm{atom}}^{\mathrm{bulk}}$(eV)&-8.941&-5.185&-6.069&-7.394&-9.305\\
LC (Å)&3.875&3.957&3.977&3.850&3.815\\
\hline
\end{tabular}
\end{table}

\begin{table}[t]
\centering
\caption{\label{tab:3}%
Parameters $\alpha$ and $\beta$ of the regression lines between the CO adsorption energy and the local surface energy (LSE) for monometallic systems, obtained using NNP and DFT methods. The values correspond to Figures~\ref{fig:2} and~\ref{fig:7}, respectively, where NNP and DFT in parentheses indicate the method used to obtain these parameters.}
\begin{tabular}{cccccc}
&Ir&Pd&Pt&Rh&Ru\\
%\mbox{Three}&\mbox{Four}&\mbox{Five}\\
\hline
$\alpha$(DFT)&-0.639&-0.446&-1.073&-0.345&-0.041\\
$\alpha$(NNP)&-0.617&-0.692&-1.131&-0.441&-0.515\\
$\beta$(DFT)&-1.782&-1.305&-1.521&-1.755&-2.012\\
$\beta$(NNP)&-1.289&-1.165&-0.983&-1.234&-1.240\\
\hline
\end{tabular}
\end{table}

\begin{figure}[]
\includegraphics{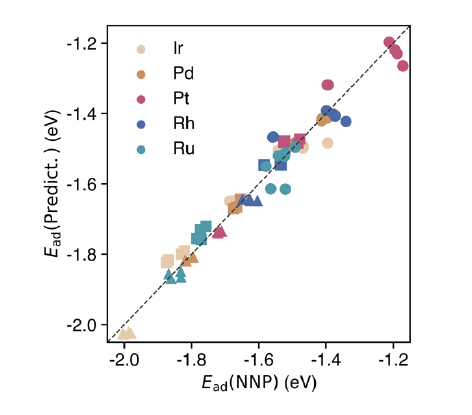}% Here is how to import EPS art
\caption{\label{fig:6}Correlation between the adsorption energies of a CO molecule on monometallic NPs calculated directly using the NNP and those predicted through regression. Circles, squares, and triangles at each datapoint represent the facet, edge, and corner CO adsorption sites, respectively.}
\end{figure}

\begin{figure}[]
\includegraphics{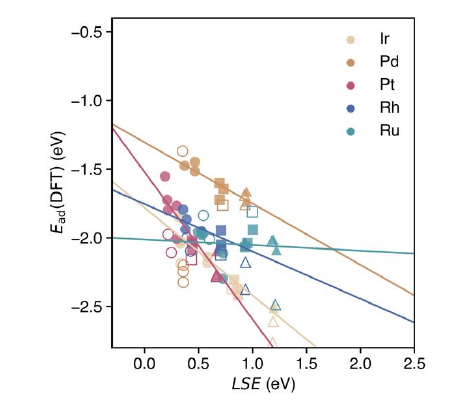}% Here is how to import EPS art
\caption{\label{fig:7}Correlation between adsorption energies of a CO molecule on monometallic NPs calculated directly via DFT and the LSEs corresponding to each adsorption site. The CO adsorption energies for the 38- and 79-atom NPs, as well as the datapoints corresponding to the hollow sites that converged on the Pd NPs, are shown as open markers, as they were not included in the linear regression data points. Circles, squares, and triangles at each datapoint represent the facet, edge, and corner CO adsorption sites, respectively.}
\end{figure}

\begin{figure*}[]
\includegraphics{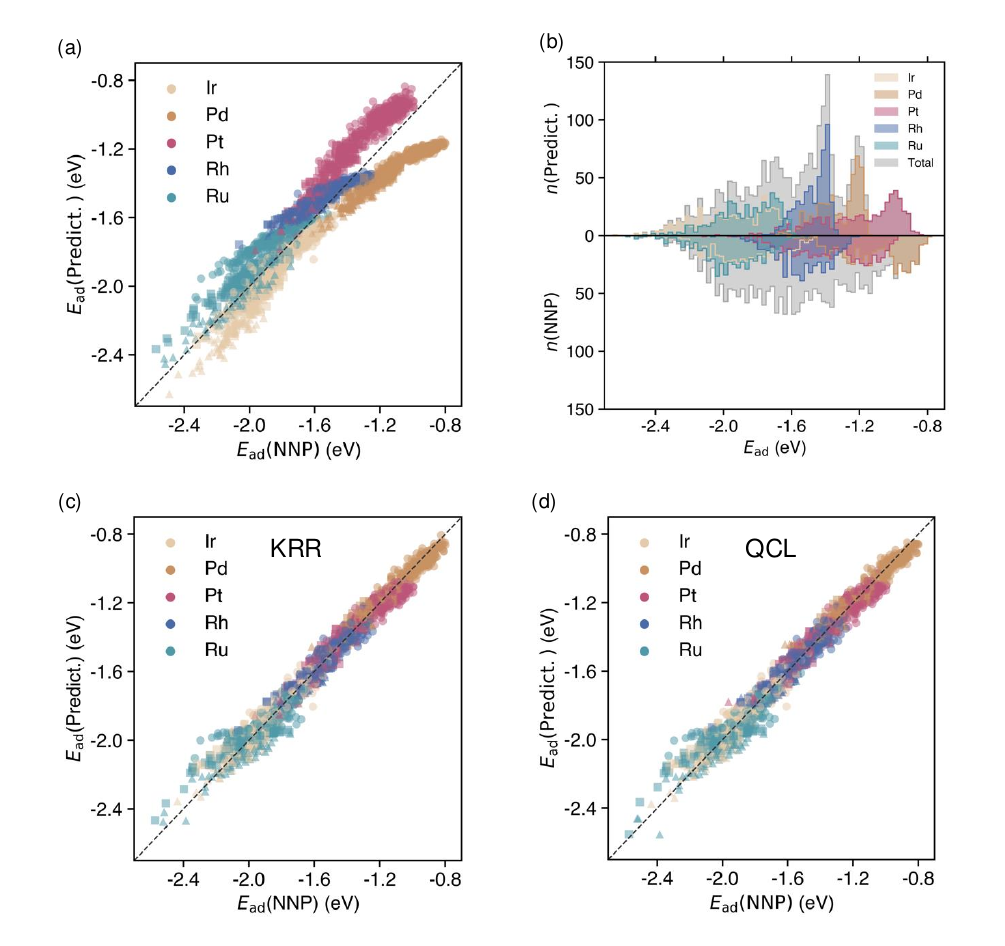}% Here is how to import EPS art
\caption{\label{fig:8}(a) Comparison and (b) correlation between the adsorption energy predicted via LSE and the adsorption energy of $\mathrm{CO/HEA}_{201}$ directly calculated using the NNP. We limit the comparison to 2440 sites, encompassing all on-top sites across the 20 structures. For each element (Ir, Pd, Pt, Rh, Ru) for 14 patterns of IrPdPtRhRu $\text{HEA}_{201}$ NPs, the energies are corrected using nonlinear regression models based on (c) KRR and (d) QCL regression. These values are plotted against values directly obtained using the NNP, which serve as the standard for accuracy.}
\end{figure*}

\section*{Data Availability}

The data and code required to reproduce the figures and tables are publicly accessible on GitHub \url{https://github.com/TShiotaSS/lse}. The datasets for this study are available on Figshare at DOI: \url{https://doi.org/10.6084/m9.figshare.26973409.v2}. They contain structural and energetic information for both HEA and monometallic NPs. The datasets also include adsorption energy data from M3GNet and VASP calculations. The provided data includes relaxed structures and DFT calculations performed at the PBE level using VASP. Additionally, key descriptors such as GCN and LSE from M3GNet are included. We modified the code to extract the atomic energies from the pretrained M3GNet model on Github at \url{https://github.com/materialsvirtuallab/m3gnet}, which can be found at \url{https://github.com/TShiotaSS/lse/tree/main/scripts/m3gnet_each_atom_energy}. The implementation for QCL regression used in this study is available at \url{https://github.com/Qulacs-Osaka/scikit-qulacs/tree/main}. 

\section*{Conflicts of interest}
There are no conflicts to declare.

\section*{Appendix}

Here, we summarize the parameters of the regression model for adsorption-energy prediction using the LSE descriptor as well as the results of the model accuracy validation. Figure~\ref{fig:6} shows the parity plot of the training data and prediction results for the LSE linear regression model presented in Figure 2. The RMSE for predicting the training data was 0.035 eV. Figure~\ref{fig:7} presents the regression results for the linear regression model between the CO adsorption energies on monometallic NPs obtained via DFT calculations and the LSE. Two datapoints corresponding to the hollow sites that converged on the facets of Pd in 38- and 79-atom monometallic NPs were excluded from the regression model. For small NPs, the entire system tended to exhibit molecular characteristics upon adsorption, making it difficult for a simple regression model to capture these effects . However, for larger NPs, a linear relationship with the LSE, similar to that observed for the NNP, was confirmed. The parameters of the regression models constructed using the adsorption energies from both the NNP and DFT are presented in Table~\ref{tab:3}. Figure~\ref{fig:8} presents a parity plot of the prediction results for the data not included in the training data for the nonlinear regression model of HEA NP adsorption energies evaluated directly using the NNP. The prediction accuracies were 0.0580 and 0.0579 eV, respectively, indicating that, as demonstrated in this study,  the regression models using KRR and QCL had comparable accuracy. Finally, the regression parameters of the GCN-based prediction model are summarized in Table~\ref{tab:4}.

\begin{table}[t]
\centering
\caption{\label{tab:4}%
Parameters $\alpha$ and $\beta$ of the regression lines between the CO adsorption energy and the generalized coordination number (GCN) for monometallic systems. The values correspond to the lines in Figures~\ref{fig:9}(a) and (b)}
\begin{tabular}{cccccc}
&Ir&Pd&Pt&Rh&Ru\\
%\mbox{Three}&\mbox{Four}&\mbox{Five}\\
\hline
$\alpha$(NNP)&0.185&0.139&0.173&0.078&0.121\\
$\beta$(NNP)&-2.766&-2.403&-1.073&-1.946&-2.371\\
\hline
\end{tabular}
\end{table}

\section*{Acknowledgements}
We thank Nobuki Inoue, Tuan Minh Do, and Ryo Watanabe for fruitful discussions.
This project was supported by funding from the MEXT Quantum Leap Flagship Program (MEXTQLEAP) through Grant No. JPMXS0120319794, and the JST COI-NEXT Program through Grant No. JPMJPF2014. The completion of this research was partially facilitated by the JSPS Grants-in-Aid for Scientific Research (KAKENHI), specifically Grant Nos. JP23H03819 and JP21K18933. We thank the Supercomputer Center, the Institute for Solid State Physics, the University of Tokyo, for allowing us to use these facilities. This work was also achieved using SQUID at the Cybermedia Center, Osaka University.

%%%END OF MAIN TEXT%%%


\balance


%%%REFERENCES%%%
\bibliography{rsc} %You need to replace "rsc" on this line with the name of your .bib file
\bibliographystyle{rsc} %the RSC's .bst file

\end{document}